\documentclass[]{emulateapj-rtx4}
\usepackage{psfig}

\shorttitle{HBO and NBO variations in GX 5-1}
\shortauthors{Sriram et al.}

\begin{document}
\title{Coupled HBO and NBO variations in the Z source GX~5-1: inner accretion disk as the 
location of QPOs}

\author{K. Sriram\altaffilmark{1}, A. R. Rao\altaffilmark{2} and C. S. Choi\altaffilmark{1}}
\altaffiltext{1}{Korea Astronomy and Space Science Institute, Daejeon 305-348, Republic of Korea
\email{astrosriram@yahoo.co.in}}
\altaffiltext{2}{Tata Institute of Fundamental Research, Mumbai 400005, India}

\begin{abstract}

The simultaneous and coupled evolution of horizontal branch oscillation (HBO) and normal
branch oscillation (NBO) in Z-type sources suggests that the production of HBO is connected to 
NBO and is caused by
changes in the physical/radiative properties of the inner accretion disk, although
there is a lack of substantial spectral evidence to support this. In this {\it Letter}, 
we present the results
of an analysis of a  RXTE observation of a Z source GX~5-1, where the 6 Hz NBO is simultaneously detected
along with a HBO at 51 Hz. 
The variations in the intensity and the associated power density spectrum indicate that the HBO
and NBO are strongly coupled, originating from the same location in the inner accretion disk. 
The absence of HBO and NBO in the lower energy bands, an increase in  the
rms amplitude with energy and a smooth transition 
among them suggest that they are produced  in the hot inner  regions of the accretion disk.  
Based on a spectral analysis, we found a signature of changing or physically modified
inner disk front during the coupled HBO and NBO evolution.
We explore the various models to explain the observed phenomenon and propose that the
NBO is affiliated to the oscillations in the thick/puffed-up inner region of the accretion disk.

\end{abstract}

\section{Introduction}

Various mechanisms have been suggested to explain the quasi-periodic oscillations (QPOs) in neutron
star low mass X-ray binaries (NS LMXBs), but the location of their production region in the accretion
disk and the connecting physical/radiative processes are not yet clarified. 
Of the two broad classification of NS LXMBs viz. atoll and Z-type sources, the latter has
a high X-ray luminosity of $\sim$ 10$^{38}$ erg s$^{-1}$ (close to the Eddington limit), which is  
secularly varying along three main branches forming a Z shape in the hardness intensity and color-color diagram. 
The associated branches from top to bottom are horizontal branch (HB), normal branch (NB) and flaring branch (FB)
and NB connects the other two (Hasinger \& van der Klis 1989; Hasinger et al. 1990). 
Broad features in the power density spectra  known as quasi-periodic oscillations 
(QPOs) are often observed along the branches. Apart from twin kHz QPOs (van der Klis 1998), 
different low and intermediate frequency QPOs have been observed in Z-type NS LMXBs named 
after their branches viz. horizontal branch oscillations (HBOs), normal branch oscillations 
(NBOs) and flaring branch oscillations (FBOs) (van der Klis 2006). 
The systematic Z-track evolution and the different oscillations are considered to be caused by 
changes in the mass accretion rate but a few studies seem to rule out any relation
to the accretion rate (Lin et al. 2009; Homan et al. 2010).

As the Z sources traverse from HB to NB, the HBO smoothly changes to NBO 
(typically a $\sim$50 Hz peak is converted into  a $\sim$6 Hz peak), though the
physical causes in the accretion disk, responsible for this variation, are unknown.  
There are two main models which generally explain the observed features of NBO. In the
radiation-hydrodynamical model (Fortner et al. 1989; Miller \& Lamb 1992), 
NBO are associated with the oscillations in the optical depth of radially inflowing hot material.
In the other model, NBO is caused as a result of Keplerian rotation of thick accretion disk,
with variations in its dynamical properties (Alpar et al. 1992). 

Many models were proposed to explain the generation of HBOs. In the
magnetospheric beat-frequency model, HBO arises due to the orbital motion at a
suitable magnetospheric radius (Alpar \& Shamam 1985; Lamb et al. 1985). The sonic-point beat
frequency model, an extension of the magnetospheric beat-frequency model, was proposed to explain
the simultaneous presence of HBOs and kHz QPOs (Miller et al. 1996, 1998).  
In the relativistic precession model (Stella \& Vietri 1999; Stella et al. 1999), HBOs and kHz QPOs are caused by the 
relativistic frame dragging and relativistic periastron precession of an  eccentric ring. In the
two oscillator model (Titarchuk \& Osherovich 1999; Titarchuk et al. 1999), 
HBO and kHz QPOs are the result of oscillations in a hot blob and the Keplerian motion of the disk.

Both NBO and HBO are observed  on numerous occasions, but 
much importance has been
given to the timing domain analysis (Cir X-1, Soleri et al. 2009; GX 340+0, Penninx et al. 1991; 
GX 17+2, Wijnands et al. 1996; Homan et al. 2002; Cyg X-2, Wijnands et al. 1997; GX 5-1, Dotani 1988). 
Yu (2007) found HBO and NBO along with twin kHz QPOs in the source Sco X-1 and suggested that 
NBO resulted from the outward disk movement but there was no systematic spectral 
study to support this idea.
In this {\it Letter}, we present the timing and spectral analysis of a RXTE observation of GX 5-1 during which
a smooth transition of HBO--NBO--HBO is observed. The spectral study indicates that the inner disk edge moves
out radially or it gets thickened or puffed-up and the properties of the disk are substantially modified 
when compared to the properties of the disk with HBO. 

GX 5-1 is the second brightest Z source (Bradt et al. 1968), located at a distance of 9.0$\pm$2.7 kpc (Christian \& Swank 1997) 
with a luminosity in the range of 6.0 -- 7.6 $\times$ 10$^{38}$ erg s$^{-1}$ (1 -- 30 keV) (Jackson et al. 2009),
and it resembles Cyg X-2 due to its HB nature (Kuulkers et al. 1994). This source exhibits
HBO and NBO along with kHz QPOs (van der Klis et al. 1985; Lewin et al. 1992; Wijnands et al. 1998) but 
FBO has not been detected yet. The kHz QPOs are not seen at NB and FB which are often observed in Sco X-1. 
The detection of radio and infrared emission provides a clue for the existence of a jet 
(Fender \& Hendry 2000; Jonker et al. 2000).      

\section{Data Analysis and Results}

RXTE observed the source GX 5-1 on 1997 July 28 spanning two satellite orbits (ObsId. 20055-01-04-00). 
We used the generic binned data ($B\_250us\_4M\_0-35\_Q$) for obtaining the power density spectrum (PDS)
and the 16 s binned data (standard 2 data) from the Proportional Counter Unit 2 (PCU2) for spectral
studies; PCU2  is the best calibrated among the PCUs (Jahoda et al. 2006). We added 0.5\% systematic errors to the
spectral data to account for the calibration uncertainties.
During this observation GX 5-1 was {\bf in} the upper NB and the intensity was nearly constant in the first
orbit (14913 $\pm$ 50 c/s) with a HBO at 49$\pm$1 Hz but the intensity varied considerably in the
second orbit, showing a dip-like feature.

The left panel in Figure 1 shows the light curve of the second orbit in the energy range 2 -- 13 keV,  
which is divided into 8 segments (a to h) for a detailed analysis. The intensity decreases to a lower state
($\sim$ 14000 c/s) in about 1000 s and subsequently recovers to the initial state ($\sim$ 15000 c/s),
similar to the variation seen in the X-ray dips of LMXBs.
The right panels in Figure 1 show the PDS in different energy bands, where the binned data are available
in four energy bands i.e. 1.94 -- 3.70 keV, 4.05 -- 5.82 keV, 6.18 -- 8.68 keV, 9.03 -- 12.99 keV and 
the event mode for the 13.36 -- 118.00 keV band.
A NBO (6.0$\pm$0.7 Hz with a rms 2.00$\pm$0.22\%) and a HBO (51.0$\pm$1.0 with a rms 2.28$\pm$0.30\%) were
found in the energy band 1.94 -- 12.99 keV.
The 6 Hz NBO and 51 Hz HBO were not present in the lower ($<$ 5.82 keV) and higher ($>$13.36 keV) 
energy bands and were prominent in the intermediate energy bands i.e. 6.18 -- 8.68 keV (7.1$\sigma$ for HBO and 7.5$\sigma$ for NBO) 
and 9.03 -- 12.99 keV (3.2$\sigma$ for HBO and 6.5$\sigma$ for NBO). 
The HBO rms increased from 2.10$\pm$0.30\% (6.18 -- 8.68 keV) to 4.94$\pm$0.76\% 
(9.03 -- 12.99 keV) and NBO rms increased from 1.92$\pm$0.23\% to 4.07$\pm$0.49\%. 

Figure 2 displays the best-fit PDS for the phases of ingress (or phase I: a+b+c),
dip (II: d+e+f), and egress (III: g+h). In phase I the HBO was detected at a significance level
of 6.4$\sigma$ (rms 2.58$\pm$0.31\%) and its strength decreased in phase II (rms 1.55$\pm$0.35\%) along with 
an appearance of NBO at 
6 Hz (4.5$\sigma$; rms 1.92$\pm$0.42) and later HBO was observed (4.2$\sigma$) in phase III. 
It is evident that the 6 Hz NBO is not present in phases I \& III.
To study the evolution of PDS during the intensity variation, we obtained the PDS of each segment
which is shown in Figure 3, along with the rms value and error. 
It is clear that the disappearance of HBO is almost balanced by the simultaneous appearance of NBO
 indicating a common physical origin behind the observed phenomenon. 

Both the eastern (a multi-color disk black body model plus a Comptonization model -- diskbb+CompTT) and
the western (a simple black body model, bbody, plus CompTT) approaches were used to fit
the spectra of the three phases after excluding
various other possible models such as a single power-law, CompTT, diskbb+power-law, bbody+power-law, etc.
The hydrogen equivalent column density was fixed at 6.6$\times$ 10$^{22}$ cm$^{-2}$ (Jackson et al. 2009).
We confirmed that the allowance of the column density to be free does not improve the fit significantly,
indicating that the observed intensity variation is
not related to absorption by dense blobs intervening the line of sight, 
as often observed in dipping X-ray sources.
The top-left panels of Figure 4 show the best-fit spectra by the diskbb+CompTT model and the top-right 
panels represent the residuals when we apply the best-fit parameters of the phase I spectrum to the spectra
from other phases.
The residuals in the phase II spectrum suggest that the accretion disk properties have significantly changed
at this phase.
The best-fit spectral parameters of the three phases are given in Table 1 for both the adopted approaches. 
The source luminosity (in 2 -- 10 keV) varied from 4.54 $\times$ 10$^{38}$ erg s$^{-1}$ (phase I)
to 4.44 $\times$ 10$^{38}$ erg s$^{-1}$ (phase II).

It is difficult to meaningfully quantify the variations of
spectral parameters because of the poor data statistics of the phase spectra, 
but residuals clearly indicate that there is a spectral difference between phase I and II.
To discern the difference, we investigated the minimum set of parameters which are sufficient and significant 
to justify the spectral change between the phases. We first fitted the phase I and II spectra (using the
eastern model) simultaneously
with the phase II parameters tied to the best-fit parameters obtained from the phase I spectrum. 
The high value of $\chi^2$/dof (=2998/104) clearly indicated that the spectrum has changed during the intensity 
variation (Table 1). We then found a decrease in the value of $\chi^2$/dof as diskbb normalization (N$_{disk}$) 
and CompTT normalization were allowed to be free, but the values of reduced $\chi{^2}$ were still
unacceptable, indicating the requirement of additional parameter variation. 
A substantial improvement of the fit was achieved by allowing the inner disk temperature (kT$_{in}$)
to be free.  After this step, any more significant improvement was not obtained even though
all the other parameters were allowed to be free. 
The F-test values and probabilities indicate that the minimum set of spectral parameters needed to vary are 
the normalizations of diskbb and CompTT and the inner disk temperature (Table 1). 
The simultaneous fit results for all the three phases are summarized in Table 1. 
The analysis of contour plot shows that 
the disk normalization which is a close measure of inner disk radius (N$_{disk}$ $\propto$ R$_{in}^2$) varied 
significantly (at $>$99\% confidence level, bottom-left panel of Figure 4). 
We applied the same method to the phase II and III spectra and
found similar results (Table 1 and bottom-right panel of Figure 4). 
The study strongly suggests that the inner disk radius varied during the simultaneous evolution of
HBO and NBO.            

\section{Discussion and Conclusion}

We have presented a detailed systematic study of the RXTE observation of GX 5-1 when the 
power of NBO and HBO were varying simultaneously (Figure 2 and Figure 3).  
Yu (2007) found NBO, HBO, and twin kHz QPOs simultaneously in the source Sco X-1 and suggested that the generation of NBO
is due to the radial outward movement of disk but this conclusion was not supported by any spectral signatures.
In this {\it Letter,} we confirm such a disk movement during the evolution of HBO--NBO--HBO transition
and propose that the associated inner disk edge has either significantly moved or puffed-up.

We found that the ratio of NBO to HBO rms is close to 2 (panel d in Figure 3) 
similar to the value of Sco X-1 (Yu 2007) and this ratio drops to 0.4 in panel g (Figure 3). 
The absence of HBO and NBO in low energy bands, relatively high amplitude of NBO in 9.03 -- 12.99 keV 
(Figure 1) and smooth transition of HBO--NBO--HBO (Figure 3) indicate that the QPOs are arising from 
a hot region in the inner accretion disk. The simultaneous spectral fit result suggests that 
the inner disk edge moves out when NBO is dominant and otherwise it moves in (Figure 4 and Table 1). 
The observed spectral variations can be explained in the framework of the radiation-hydrodynamical
model: the disk material leaves the inner edge and falls on to the neutron star
surface covered with a hot corona. The outward radiative pressure halts or pushes back the inner disk
edge, which can cause the relative increase of the observed disk normalization. 
Alpar et al. (1992) pointed out that the rotation of a thick disk in the inner region could
cause the observed NBOs. In this scenario, however, the twin kHz QPOs should not be seen because of the slow
dynamical movement but are often observed, e.g., in Sco X-1 (where kHz QPO is also
observed at normal and flaring branches). From the above discussion, we suggest that the 
oscillation frequency at the inner edge (at 51 Hz) is abated by the outward radiation pressure 
and results in a 6 Hz NBO. Since the vertical height and surface density of the disk are 
expected to increase during the process, the observed increase in the disk normalization 
could be explained.

In the two oscillator model, the perpendicular mode ($\nu$$_{L}$) of oscillating hot blob in 
magnetosphere generates the HBO (Titarchuk et al. 1999) and if the mass of
the blob increases, frequency of that mode would shift to a lower frequency not much affecting the 
X-ray continuum. We speculate that whatever may be the oscillating mechanism the observed smooth 
transition of HBO to NBO could be due to an increase in the participating mass in the oscillations. 
Our spectral results suggest that when the disk inner front radius which is the probable site for the 
transition boundary is increased, it gets thickened or puffed-up due both to the outward radiation
pressure and to the accumulation of matter coming from the outer region of the disk.
In this scenario, it is quite possible that the structure of the blob would be altered significantly
and the quasi-coherent perpendicular motion of the blob might have disturbed/randomized to produce
the broad 6 Hz NBO instead of 51 Hz HBO. 
The other possible scenario could be that the Keplerian motion of the disk became super-Keplerian which affects the
oscillations (Titarchuk et al. 1998) in the vertical structure of the disk and have respective dependency
on the oscillation frequencies ($\nu_{v}$ (viscous oscillations) $\propto$ L$^{-1}$ (thickness)
and $\nu_{b}$ (break frequency) $\propto$ L$^{-2}$, Titarchuk \& Osherovich 1999). The NBOs are
always found to be $<$ 20 Hz which implies that these flows would steeply affect these oscillations.

Casella et al. (2005) associated HBO, NBO and FBO to C, B and A type QPOs in black hole binaries (BHBs). 
In general type B, A QPOs and their transitions are often observed during the  
very high (VH) / steep power-law (SPL) state (McClintock \& Remillard 2004) or intermediate state (IM) (Belloni 2010). 
In this state, the accretion disk is considered to contain a compact corona (a low electron temperature with a high optical 
depth) along with a marginally truncated high temperature Keplerian disk
(Done \& Kubota 2006; Sriram et al. 2007, 2010). A broad QPO at 6 Hz is often observed in a few BHBs e.g. 
XTE J1550-564 (Homan et al. 2001), XTE J1859+226 (Casella et al. 2004), GX 339-4 (Nespoli et al. 2003; Belloni et al. 2005). 
Similar transient variation from type B to A QPO was also observed in H1743-322 (Homan et al. 2005) and GRS 1915+105 (Soleri et al. 2008). 
It was found that transient changes in QPOs are often associated with source intensity variations. 
The 6 Hz NBO is observed only during a characteristic transition in both Z-type sources and BHBs, which indicates
that a common physical process is occurring at the inner region of the disk in both the sources.

In a few atoll sources, similar QPOs varying between 5 and 14 Hz with a relatively high coherence 
were reported (Wijnands et al. 1999; Wijnands \& van der Klis 1999; Belloni et al. 2004). 
These detections and a low mass accretion rate in atoll sources pose a distinct challenge to the 
models explaining the 6 Hz NBO in Z sources. It was found that in atoll sources and BHBs 
these features are produced at the peak of their
outburst where the inner disk edge is most probably located close to the compact object. We propose that 
the mass accretion rate is not the important criterion to produce NBO features, instead it is the 
location of the inner disk which drives the NBO. If the disk edge is close to the compact object, the chances of getting
affected by the outward radiation pressure is high which would generate or disturb the oscillation
in the inner region of the disk. Similar studies concentrating on the correlated spectral and temporal variation
in LMXBs would be helpful to get a clear picture of the accretion disk during the presence of NBO/NBO-like features.

\acknowledgements 
We thank the anonymous referee for the useful comments. 
This research has made use of data obtained through HEASARC
online service, provided by NASA/GSFC, in support of the
NASA High Energy Astrophysics Programs.

\clearpage
\begin{figure}

\begin{center}$
\begin{array}{cc}
\includegraphics[width=12.0cm,height=8.5cm,angle=270]{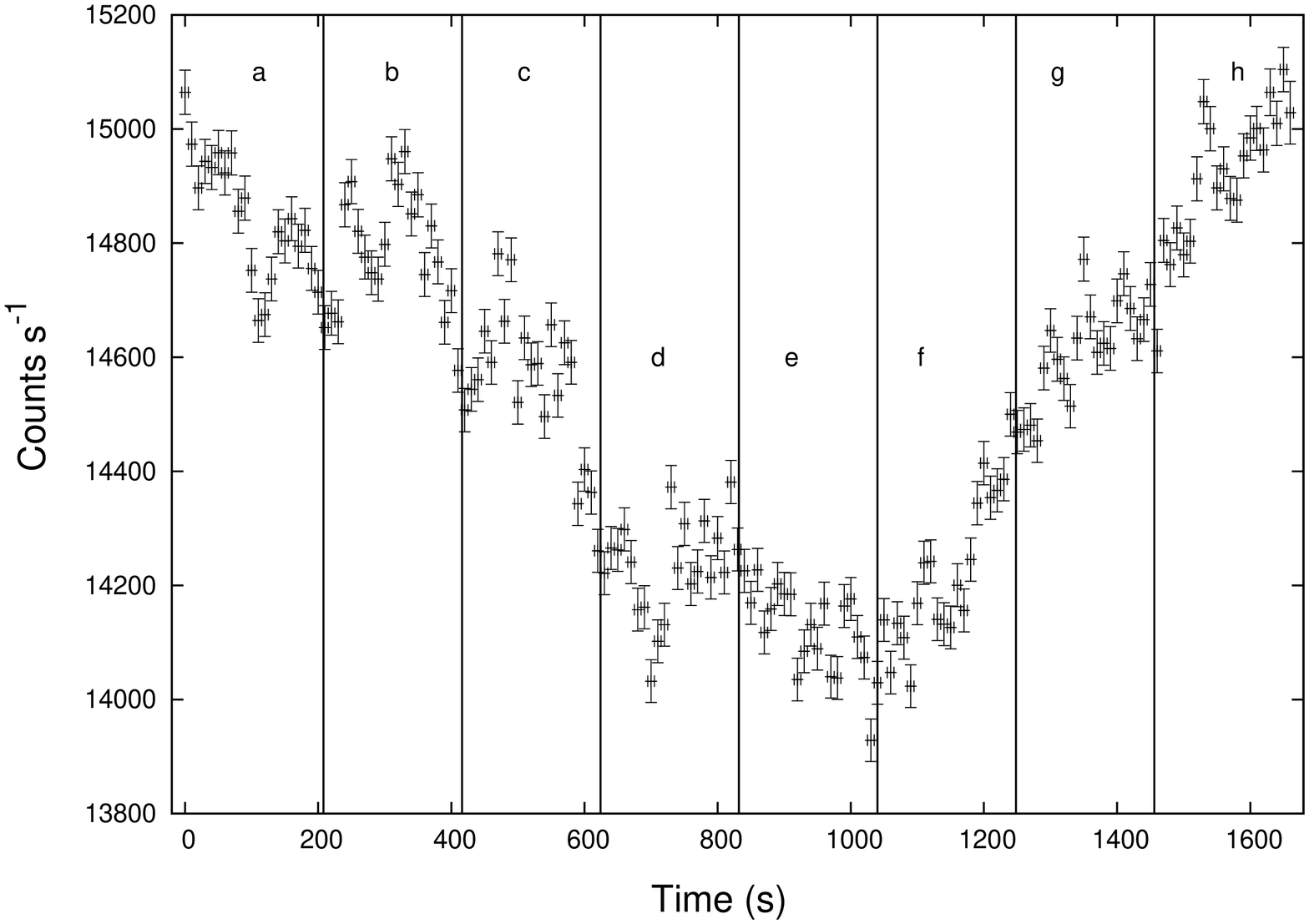} &
\includegraphics[width=12.0cm,height=8.5cm,angle=270]{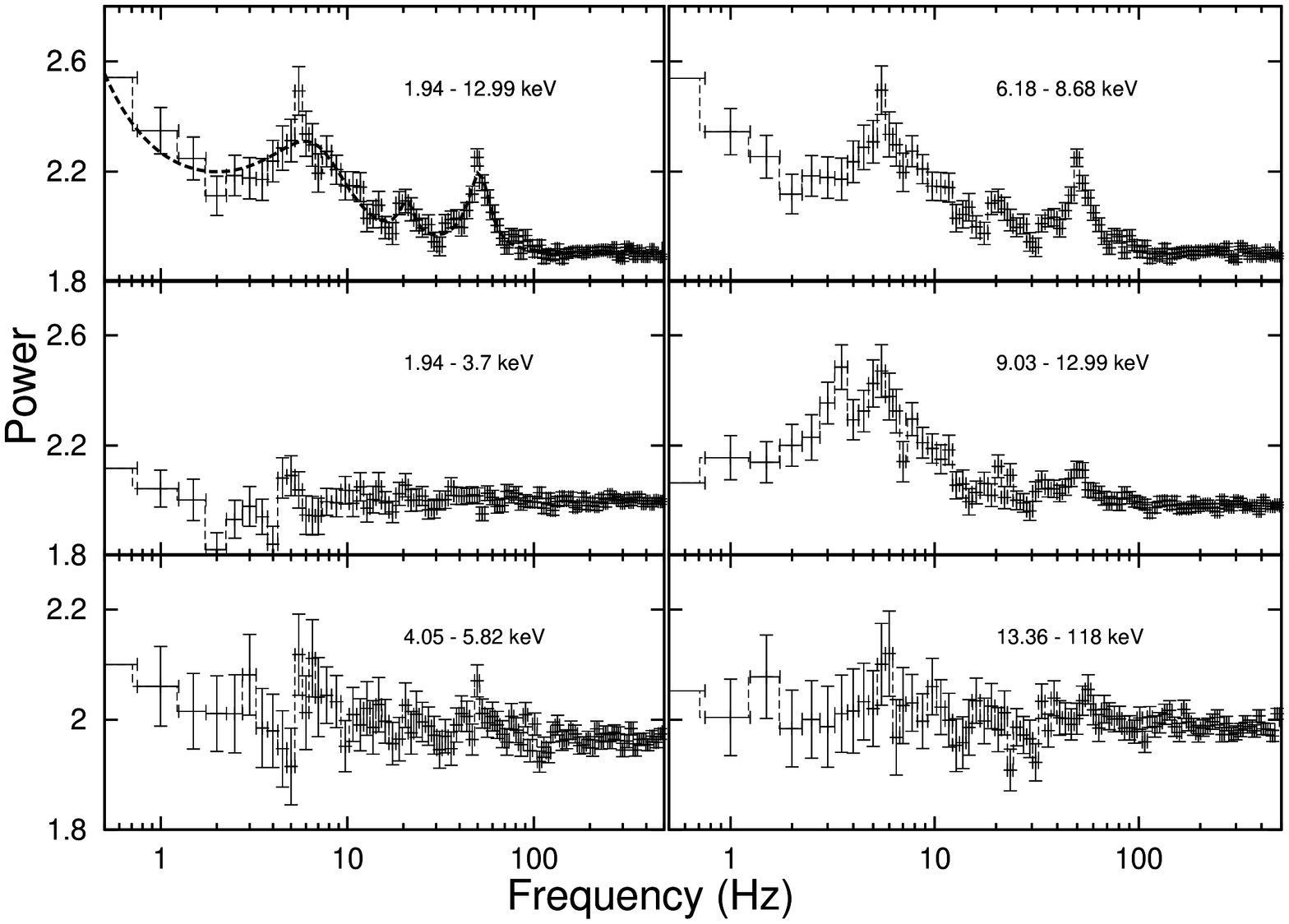} \\

\end{array}$
\end{center}
\caption{Left: The RXTE/PCA 2 -- 13 keV light curve (10 s binned) of GX 5-1 for ObsId. 20055-01-04-00. 
For detailed timing and spectral analysis, the light curve is divided into 8 segments (a to h). 
Right: The power density spectrum (PDS) of the light curve in different energy bands.
The PDS in 1.94 -- 12.99 keV is fitted by a model of a power-law with two Lorentzians. }
\end{figure}

\clearpage
\begin{figure}

\begin{center}$
\begin{array}{cc}
\includegraphics[width=15.0cm,height=15.0cm,angle=270]{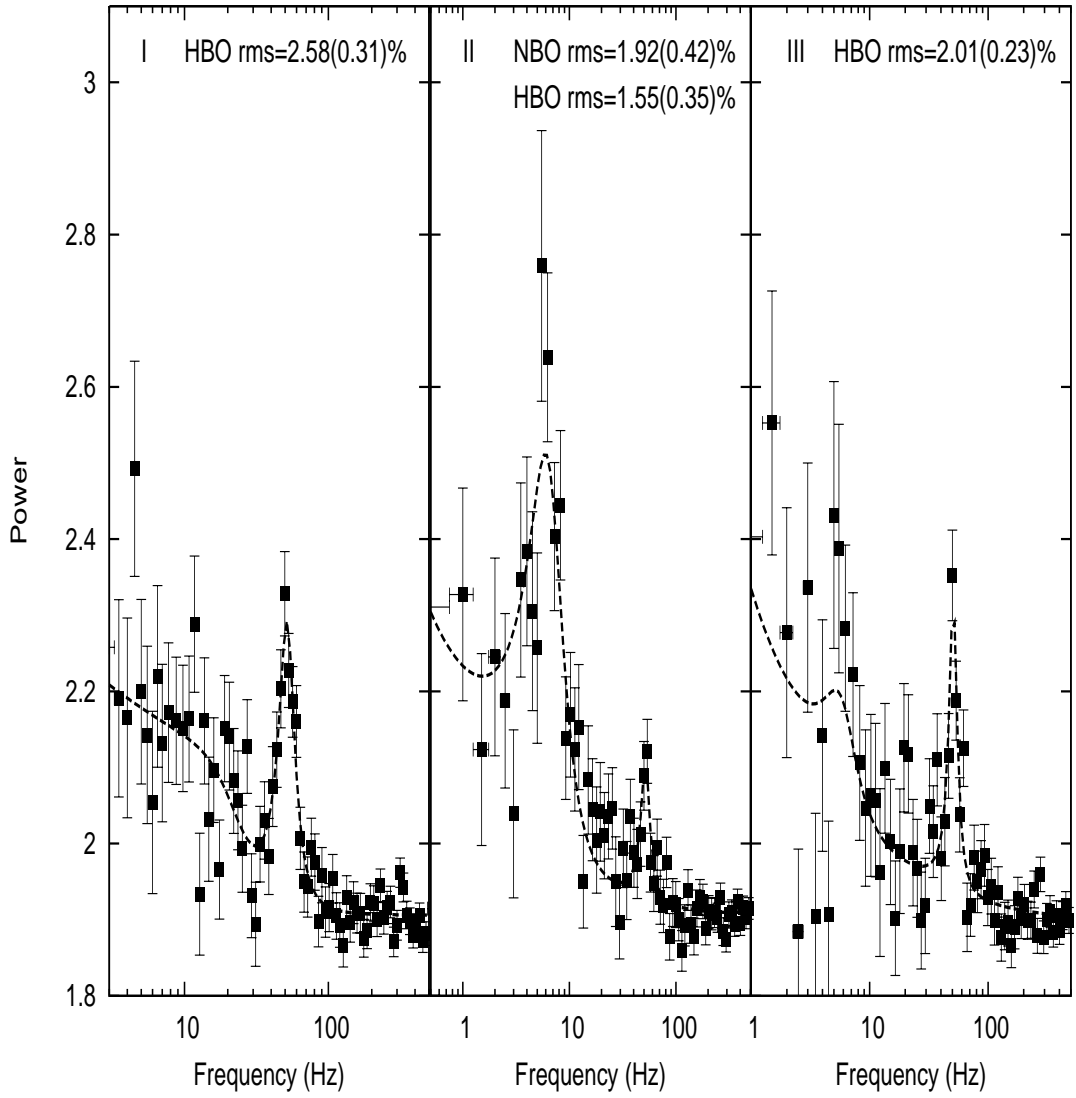}&

\end{array}$
\end{center}
\caption{The best-fit (dashed lines) PDS for the ingress (or phase I: a+b+c), the dip
(II: d+e+f), and the egress (III: g+h) phase of the light curve. 
The Poisson level is not subtracted. The HBO and NBO rms and the errors (in brackets) are 
shown.
  }
\end{figure}

\clearpage
\begin{figure}
\begin{center}$
\begin{array}{cc}
\includegraphics[width=15.0cm,height=15.0cm,angle=270]{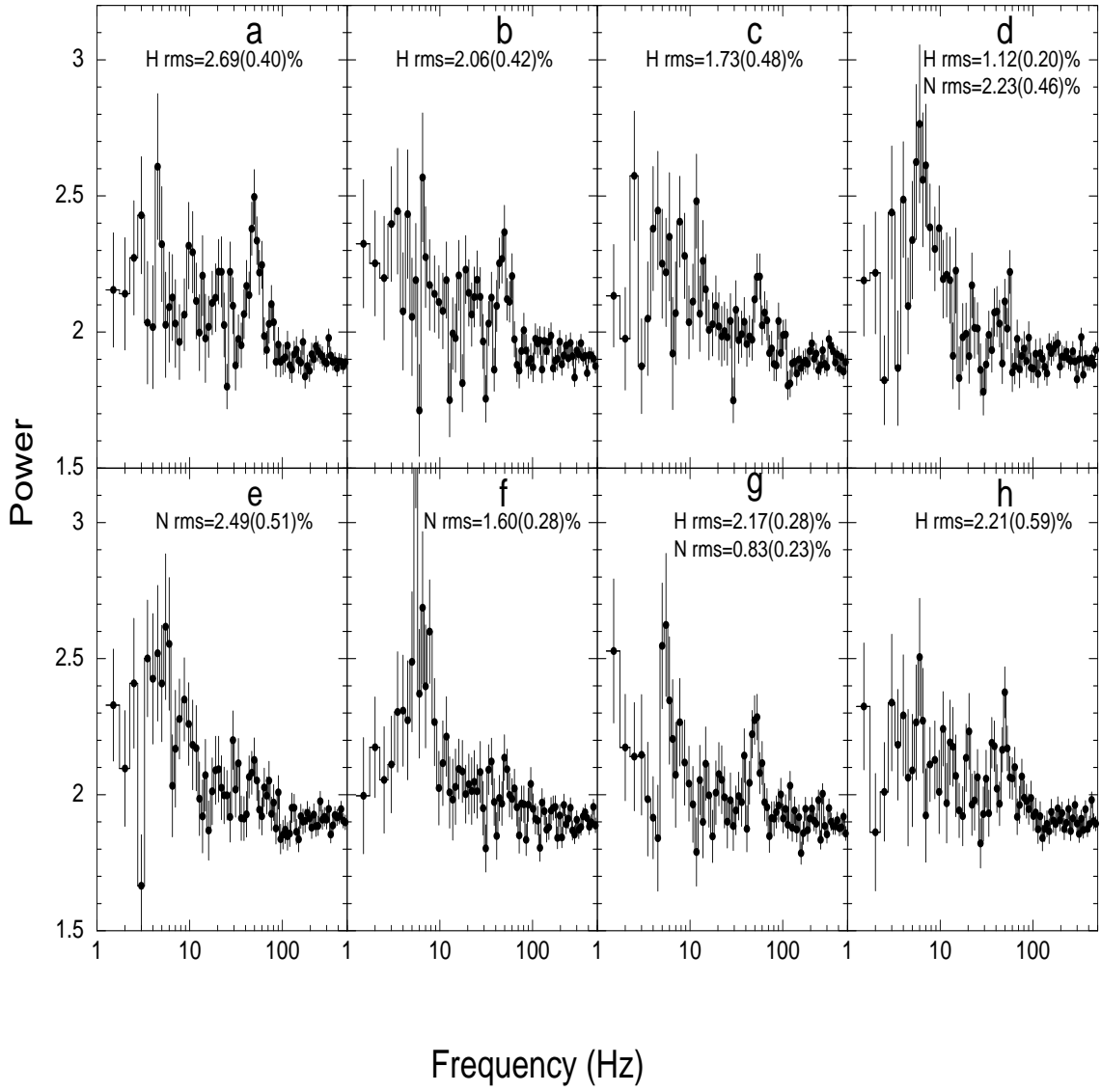} &

\end{array}$
\end{center}
\caption{The PDSs of the 8 segments showing the temporal evolution of the coupled
appearance and disappearance of HBO (H) and NBO (N). The segment label and rms powers in QPOs 
(errors given in brackets)
are given in each corresponding panel.}
\end{figure}

\clearpage
%\begin{sidewaysfigure}[b]
\begin{figure}[b]
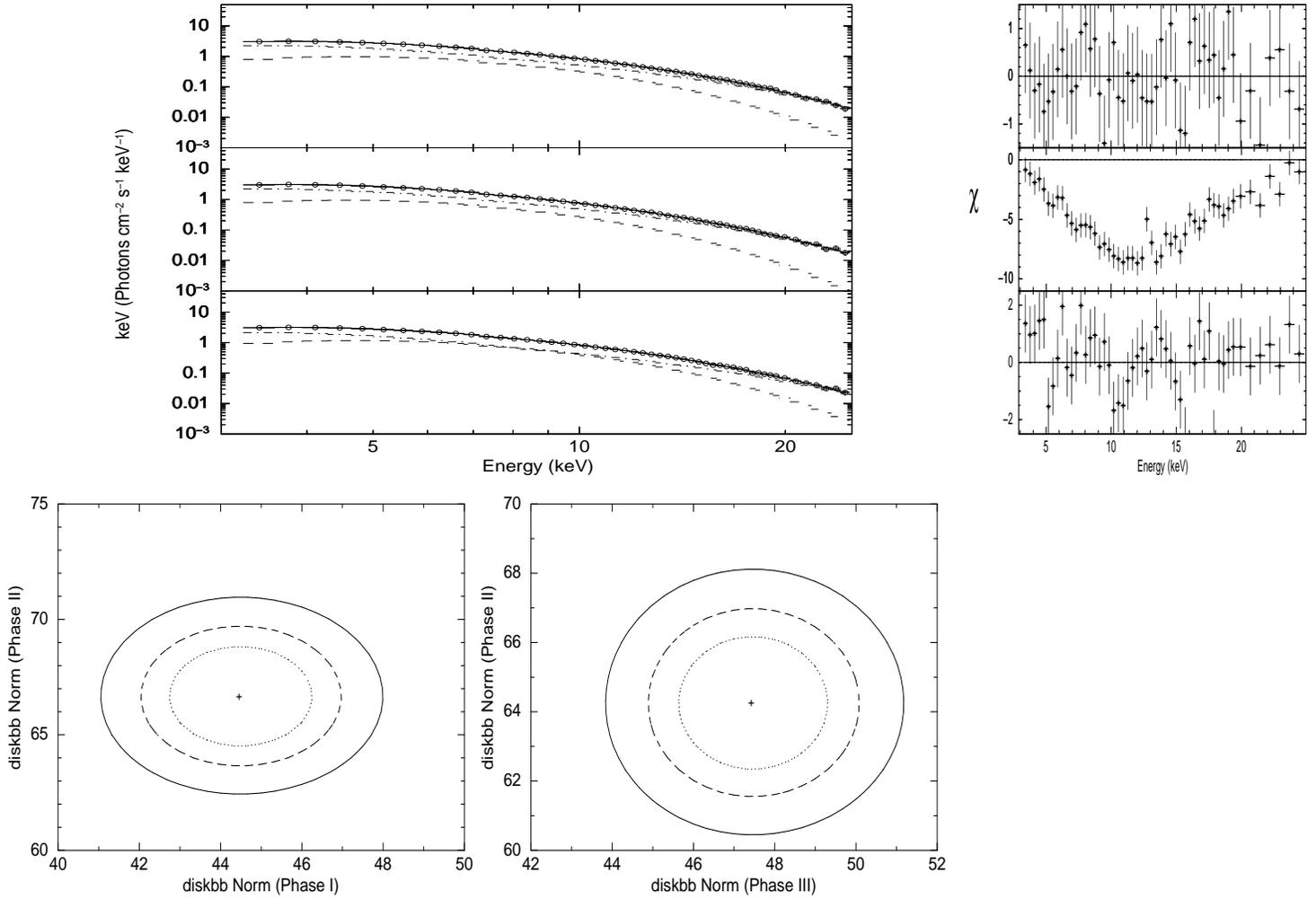

\begin{center}$
\begin{array}{cc}
\includegraphics[width=7.0cm,height=11.0cm,angle=270]{fig4.1.ps}&
\includegraphics[width=7.0cm,height=5.0cm,angle=270]{fig4.2.ps} \\
\includegraphics[width=6.0cm,height=7.0cm,angle=270]{fig4.3.ps} 
\includegraphics[width=6.0cm,height=7.0cm,angle=270]{fig4.4.ps} 
\end{array}$
\end{center}

\caption{
Top (left): The unfolded spectra for phases I (top), II (middle), and III (bottom),
using the diskbb+CompTT model are shown. Top (right): Residuals resulting from modeling the spectra of all phases
using the best-fit parameters for phase I spectrum. Bottom (left and right): Significance contour plots
between diskbb normalization (N$_{disk}$) of phase I and phase II, phase III and phase II
obtained during the simultaneous fit procedure
(see text). Contours show the 68\%, 90\%, 99\% confidence levels. Plots clearly suggests that N$_{disk}$
is different for the three phases at more than 99\% confidence level.}
%\end{sidewaysfigure}
\end{figure}

\clearpage
\setcounter{table}{0}
%\begin{table}
\begin{table}
\scriptsize
\begin{minipage}[t]{\columnwidth}
\caption{Best-fit spectral parameters for the three phase spectra. 
The corresponding subscripts dbb \& bb are diskbb \& bbody model. 
Errors are quoted at a 90\% confidence level. The unabsorbed flux in 3.0 -- 25.0 keV 
is given in units of 10$^{-8}$ ergs cm$^{-2}$ s$^{-1}$. 
The simultaneous fit results between phases I and II and phases II and III are shown. The best-fit parameter values 
from the simultaneous fit to the three phase spectra are also listed.}
\label{tab1}
\centering
\begin{tabular}{ccccccccc}
\hline
%Parameters&\multicolumn{2}{c}{ObsId 20010-01-01-00} \\
\hline
Parameters&\multicolumn{2}{c}{I}&\multicolumn{2}{c}{II}&\multicolumn{2}{c}{III}\\
\hline
&I$_{dbb}$&I$_{bb}$&II$_{dbb}$&II$_{bb}$&III$_{dbb}$&III$_{bb}$\\
 %&diskbb&bb&diskbb&bb\\
\hline
\hline

$kT_{in}$ (keV)\footnote{Inner disk temperature of the diskbb model.}&2.13$\pm$0.05 &-&2.00$\pm$0.04&-&2.17$\pm$0.05\\
$N_{disk}$\footnote{Normalization of the diskbb model.}&45$\pm$5 &-&54$\pm$6&-&47$\pm$5\\

$kT_{bb}$ (keV)\footnote{Temperature of the single temperature black body model.}&-&1.51$\pm$0.06 &- &1.43$\pm$0.07&-&1.55$\pm$0.03\\
$N_{bb}$\footnote{Normalization of the black body model.}& -&0.079$\pm$0.016 &-&0.08$\pm$0.015&-&0.10$\pm$0.002\\
kT$_{e}$\footnote{Electron temperature.}(keV)& 3.21$\pm$0.05&2.92$\pm$0.15&3.13$\pm$0.05&2.86$\pm$0.14&3.53$\pm$0.07&3.14$\pm$0.01\\
$\tau$\footnote{Optical depth of the Compton cloud.}& 8.17$\pm$1.16&9.06$\pm$0.68&8.21$\pm$1.24&8.98$\pm$1.30&7.33$\pm$0.85&8.31$\pm$0.80  \\

 Total Flux$_{(diskbb+CompTT)}$ & 3.63& - &3.53&-&3.63&-\\
%(diskbb+CompTT)&&&&&&\\
%Diskbb flux& 1.12&-&1.09&-&1.32&-\\
%CompTT flux&2.50&-&2.43&-&2.31\\

 Total Flux$_{(bbody+CompTT)}$&-&3.74&-&3.57&-&3.75\\
%(bbody+CompTT)&&&&&&\\
%Black Body flux&-&0.55&-&0.54&-&0.71\\
%CompTT flux&-&3.19&-&3.03&-&3.04\\

$\chi^{2}$/dof&44/46 & 39/46&43/46&44/46 &36/46 &37/46\\
\hline
\hline
&Simultaneous Fit Result&&\\
&Phase I and II spectra&&\\
\hline
\hline
Parameters&$\chi^{2}$&dof&F-Stat Value&Probability&&&\\
\hline
All tied&2998&104&--&--&&&\\
Normalization's ($N_{disk}$, $N_{CompTT}$) free&899&100&58&2.52$\times$10$^{-25}$&&&\\
kT$_{in}$ along with Norms free&105&98&370&2.01$\times$10$^{-46}$&&&\\
kT$_{e}$, kT$_{in}$ along with Norms free&92&96&6.7&1.72$\times$10$^{-2}$&&&\\
All free&91&92&0.25&0.91&&&\\
\hline
\hline
&Simultaneous Fit Result&&\\
&Phase II and III spectra&&\\
\hline
\hline
All tied&2993&104&--&--&&&\\
Normalization's ($N_{disk}$, $N_{CompTT}$) free&778&100&71&2.21$\times$10$^{-28}$&&&\\
kT$_{in}$ along with Norms free&100&98&332&2.19$\times$10$^{-44}$&&&\\
kT$_{e}$, kT$_{in}$ along with Norms free&86&96&7.8&7.17$\times$10$^{-3}$&&&\\
All free&85&92&0.27&0.89&&&\\
\hline
\hline
&Simultaneous Fit Result&&\\
&Phase I, II, and III spectra&&\\
\hline
\hline
Parameters&I&II&III\\
\hline
kT$_{in}$&2.13$\pm$ 0.01& 1.96$\pm$0.01 &2.12$\pm$0.01\\
N$_{disk}$&44$\pm$ 2 &66$\pm$3  &45$\pm$2\\
%kT$_{o}$\footnote{Input soft photon temperature in CompTT model.} (keV)&0.31$\pm$ 0.06&0.39$\pm$0.04& 0.36$\pm$0.07\\
CompTT Normalization&8.95$\pm$3.28&6.13$\pm$2.86&7.54$\pm$3.29\\
\hline
\hline

\hline
\hline
\end{tabular}
\end{minipage}
\end{table}
%\end{table}

%%%%%%%%%%%%%%%%%%%%%%%%%%%%
\end{document}